# A hidden advantage of van der Waals materials for overcoming limitations in photonic integrated circuitry

*Andrey A. Vyshnevyy,[1] Georgy A. Ermolaev,[1] Dmitriy V. Grudinin,[1] Kirill V. Voronin,[2] Ivan Kharichkin,[1] Arslan Mazitov,[3] Ivan A. Kruglov,[1] Dmitry I. Yakubovsky,[1] Prabhash Mishra,[4] Roman V. Kirtaev,[1] Aleksey V. Arsenin,[1,5] Kostya S. Novoselov,[6,7,8] Luis Martin-Moreno,[9,10] and Valentyn S. Volkov[1]*

*[1]Emerging Technologies Research Center, XPANCEO, Dubai Investment Park 1, Dubai, United Arab Emirates*

*[2]Donostia International Physics Center (DIPC), Donostia/San Sebastián 20018, Spain*

*[3]Institute of Materials, École Polytechnique Fédérale de Lausanne, 1015 Lausanne, Switzerland*

*[4]Centre for Nanoscience and Nanotechnology, Jamia Millia Islamia (Central University), Jamia Nagar, New Delhi, 110025, India*

*[5]Laboratory of Advanced Functional Materials, Yerevan State University, Yerevan 0025, Armenia*

*[6]National Graphene Institute, University of Manchester, Manchester M13 9PL, United Kingdom*

*[7]Institute for Functional Intelligent Materials, National University of Singapore, Singapore 117574, Singapore*

*[8]Chongqing 2D Materials Institute, Chongqing 400714, China*

*[9]Instituto de Nanociencia y Materiales de Aragón (INMA), CSIC-Universidad de Zaragoza, 50009 Zaragoza, Spain*

*[10]Departamento de Física Aplicada, Facultad de Ciencias, Universidad de Zaragoza, 50009 Zaragoza, Spain*

*e-mail: vsv@xpanceo.com*

**ABSTRACT**

**With the advance of on-chip nanophotonics, there is a high demand for high refractive index, low-loss materials. Currently, this technology is dominated by silicon, but van der Waals (vdW) materials with high refractive index can offer a very advanced alternative. Still, up to now it was not clear if the optical anisotropy perpendicular to the layers might be a hindering factor for the development of vdW nanophotonics. Here, we studied $WS_2$-based waveguides in terms of their optical properties and, particularly, in terms of possible crosstalk distance. Surprisingly, we discovered that the low refractive index in the direction perpendicular to the atomic layers improves the characteristics of such devices, mainly due to expanding the range of parameters at which single-mode propagation can be achieved. Thus, using anisotropic materials offers new opportunities and novel control knobs when designing the nanophotonic devices.**

**Introduction**

The isolation of graphene[1], followed by the exploration of the ever-growing family of 2D materials[2,3], has revolutionized modern physics. Their van der Waals (vdW) nature makes possible the creation of heterostructures[4,5], a new form of artificial materials with on-demand characteristics[6]. More importantly, 2D crystals are compatible with existing 3D material platforms owing to the developed transfer techniques[7–9], which opens a broad range of applications. They include flexible nanoelectronics[10,11], tattoo medicine[12,13], photodetection[14,15], topological sensing[16,17], and thermal management[18–20] to name just a few.

Concerning next-generation nanophotonics, 2D materials have been employed for light generation, detection and manipulation. Thanks to the ultrafast response of monolayers, they are promising as parts of integrated modulators[21,22], nanoscale light emitters[23,24], and on-chip photodetectors[25,26].

Besides monolayers, current research also focuses on bulk films and structures of vdW materials. Similarly to monolayers, they often show strong optical responses with distinct excitonic resonances resulting in high refractive index[27] reaching 4. It enables compact nanoresonators[28], strong light-matter coupling[29], and anapole-exciton polaritons[30]. Also, they were recently recognized as a self-sufficient material platform that may complement or even replace silicon in integrated photonic circuits[31]. Besides, with slight modifications, silicon fabrication technologies allow one to create nanostructures from layered materials with an almost atomic precision,[20,32,33] thereby placing industrial incorporation of vdW materials within reach.

Another key optical characteristic of vdW materials is the giant negative optical anisotropy, which was recently determined by several groups[34–36] and is intimately related to the record-breaking refractive indices[37]. It might detrimentally impact the otherwise superior properties of vdW material-based nanophotonic components because of relatively low out-of-plane component of refractive index $n \sim 2.5$. Although in claddings, such an anisotropy can be harnessed to achieve subdiffractional guiding,[34,38,39] its impact in vdW material-based integrated photonic components is yet to be understood.

In this work, we reveal nontrivial influence of the optical anisotropy in integrated photonics. Counterintuitively, it improves the practically relevant characteristics of vdW material waveguides, in particular, leading to the reduction in the waveguide cross-talk, thereby pushing the limits of integration beyond isotropic nanophotonics.

**Results**

**Optical response of TMDCs**

To investigate TMDC features in integrated nanophotonics, we focus on WS$_2$ since it has one of the highest stability and optical bandgap among traditional TMDCs.[30,40] We commence our analysis with optical constants determination through ellipsometric measurements (Figure 1 and Supplementary Note 1).[34] They result in etalon optical constants, as confirmed by a good agreement between experimental and theoretical WS$_2$ dielectric tensors in Figure 1a. Similar to MoS$_2$,[34] WS$_2$ exhibits giant optical anisotropy (Figure 1b) and high-refractive index (Figure 1c). We recorded planar waveguide modes (Figure 1d-e and Supplementary Note 2) in WS$_2$ flakes via scattering-type scanning near-field optical microscope (s-SNOM) to confirm these observations. Fourier mode analysis (Figure 1f) reveals a fundamental mode whose dispersion perfectly fits transfer matrix calculations, presented in Figure 1g, thereby unambiguously validating dielectric function in Figure 1a.

Interestingly, in terms of refractive index and optical bandgap, WS$_2$ is beneficial for all-dielectric nanophotonics compared to classical highly refractive materials, as seen in Figure 1c. More importantly, if we consider other layered materials, such as MoS$_2$ and hBN, one can notice that they form a next-generation optical platform (Figure 1c). Nevertheless, active research in all-dielectric nanophotonics establishes high-refractive index materials, which cover the whole spectral range up to 4 eV, as illustrated in Figure 1c. In contrast, layered materials have a spectral gap between 2 and 4 eV, which requires further search for layered materials with superior optical response for these wavelengths.

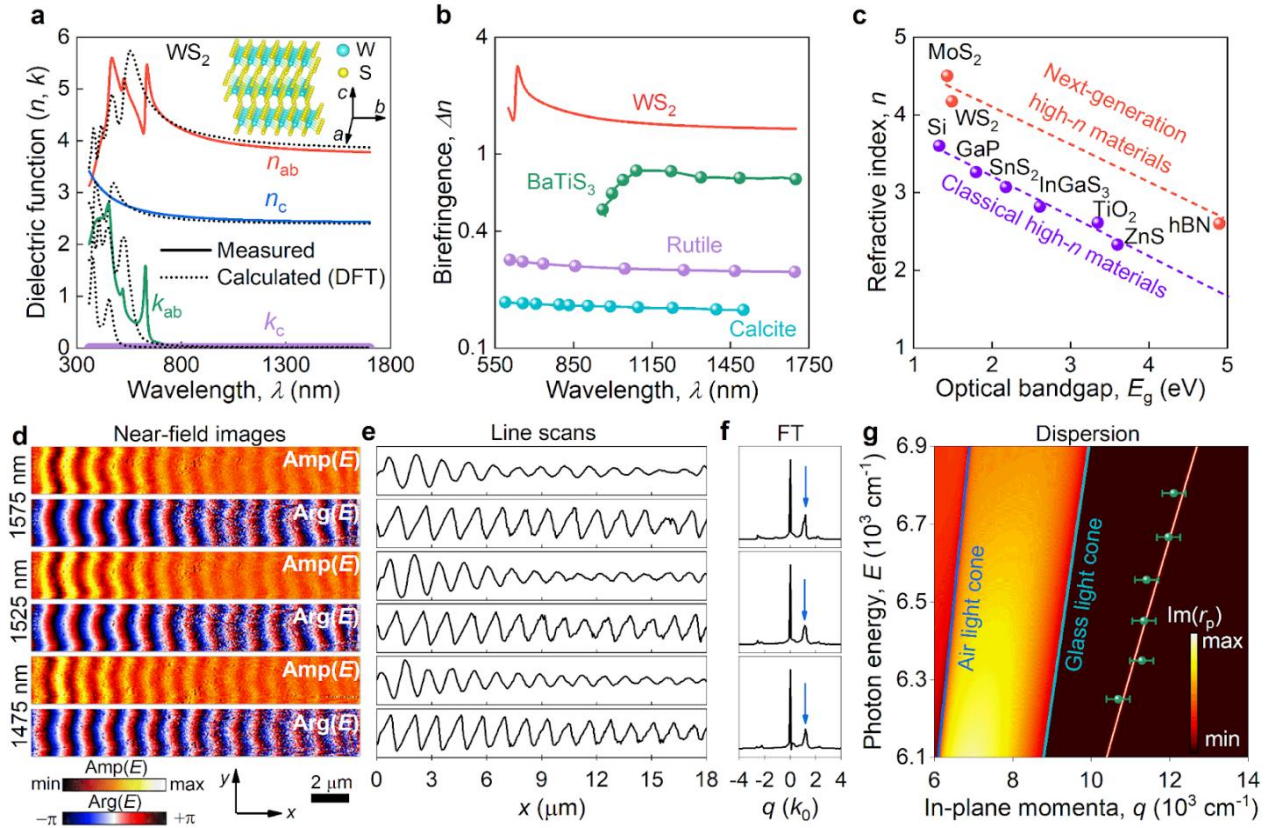

**Figure 1. Anisotropic optical properties of WS$_2$. a,** Dispersion of in- ($n_{ab}$ and $k_{ab}$) and out-of-plane ($n_c$ and $k_c$) components of refractive index and extinction coefficient (solid lines) and their comparison with the results of *ab initio* calculations (dashed lines). **b,** Birefringence of WS$_2$ and its comparison with other well-known birefringent materials. **c,** Comparison of the maximum in-plane refractive index of WS$_2$ in the transparency window with other established highly refractive materials. **d,** Amplitude and phase maps of a WS$_2$ flake measured by s-SNOM. **e,f,** line

scans in panel **e** and their Fourier transforms in panel **f** measured at varying wavelengths. **g,** Measured dispersion of the planar waveguide mode (green dots) and its comparison with the dispersion (solid orange line), calculated in transfer matrix framework using the optical constants shown in panel **a**.

**Impact of anisotropy on integrated circuits**

Next, we theoretically study how the high refractive index and optical anisotropy of $WS_2$ can be harnessed to improve photonic integrated circuits (PICs). Waveguides are the backbone of the PICs, and their network's density and complexity determines the PIC's performance. However, the electric field in dielectric waveguides permeates into surrounding space and cladding, which gives rise to a parasitic crosstalk (Figure 2a), whose strength is characterized by the crosstalk distance $L_{ct}$. To avoid bit errors, the crosstalk distance should significantly exceed the characteristic size of a PIC (usually a few mm). This restriction eventually determines how close independent waveguides can be.

Setting the waveguide width to 500 nm, wavelength $\lambda$ to 1550 nm, and varying the waveguide height, we evaluate the effective mode index of the fundamental mode $n_{eff}$, crosstalk distance $L_{ct}$ and mode area, with the latter being a widely accepted measure of the field confinement. Contrary to expectations, parameters maximizing the crosstalk distance do not minimize the mode area. Instead, both the crosstalk distance and the mode area increase as the waveguide height increases from 120 to 300 nm (Figure 2b). At the same time, we notice that the crosstalk distance correlates with the effective index of the fundamental waveguide mode.

To explain the observed behavior, we examine the electric field distribution in a single waveguide. Figure 2c shows that more than an order of magnitude increase in the crosstalk distance is caused by the decrease in the evanescent field strength created by the left waveguide at the location of the right waveguide. The effective mode index $n_{eff}$ determines the imaginary wavevector of the evanescent field through $\kappa \approx \frac{2\pi}{\lambda}\sqrt{n_{eff}^2 - n_{glass}^2}$. Thus, the evanescent tails of modes with a higher effective index decay faster in the surrounding waveguide space. By contrast, the mode area is much less dependent on the evanescent field, which results in lack of correlation between the mode area and the crosstalk distance.

Figure 2d illustrates the field distribution in the fundamental waveguide mode and the waveguide crosstalk. It shows the calculated field distribution at a distance of $L$ = 0.15 mm from the inlet, assuming that only the left waveguide carries the signal at the inlet. For a 120-nm-high waveguide, almost half of the signal power crosses over to the right waveguide ($L_{ct} \approx 2L$), while there is no observable crosstalk in 300-nm-high waveguides ($L_{ct} \gg L$).

To optimize waveguide dimensions, we evaluate crosstalk distance as a function of the dimensions of the waveguide core (Figure 2e). At a fixed height, with the increase in width, the crosstalk distance reaches the maximum and then starts decreasing. This is a result of two opposing factors. On the one hand, at greater widths, the effective index and, consequently, $\kappa$ are higher, which entails more rapid evanescent decay. On the other hand, with the increase in the width, the distance between the nearest points of waveguide cores gets smaller, preventing the accumulation of evanescent decay. By contrast, at a fixed width, the crosstalk distance monotonically increases with the increase of height, following the trend established in Figure 2b. However, PICs are usually designed to operate in the single-mode regime,[41] which imposes limitations on the waveguide core width and height, shown by the white line in Figure 2e. We find that the maximum crosstalk distance of 4.7 mm is achieved with a core width and height of 440 nm and 240 nm, correspondingly. These sizes exceed width and height where the minimum mode area is achieved, which again confirms that a waveguide with the smallest mode area may not be the most suitable for PIC.

Finally, we investigate the influence of optical anisotropy on waveguide characteristics. We similarly study the crosstalk distance, assuming that the core material is isotropic with a refractive index equal to the in-plane refractive index of $WS_2$ (Figure 2f). We note that for the same width and height, the crosstalk distance of the isotropic-core waveguide is only a few percent larger than that of the anisotropic one. This can be explained by the change in the effective mode index on the order of $10^{-4}$, which is surprisingly small in light of the giant optical anisotropy of $WS_2$. Such a tiny difference is due to the small y-component of the fundamental mode electric field, as confirmed by Figure 2d, where the electric field value is discontinuous across vertical sidewalls of waveguides and continuous across top and bottom surfaces. However, such an argument no longer applies to the higher-order modes. As a result, cut-off curves for higher-order modes of anisotropic-core waveguide lie much higher than those of isotropic one. Consequently, optimized waveguide dimensions and the waveguide crosstalk distance are significantly smaller for the isotropic core (1.5 mm) than for anisotropic one (4.7 mm).

To gain further insight into the influence of anisotropic optical constants, we determine the optimal crosstalk distance at different $n_c$'s, starting from $n_c = 1$ and ending at $n_c = n_{ab}$. Moreover, we add other layered materials for comparison, namely, $MoS_2$ and $MoSe_2$. In all cases, the optimized crosstalk distance decreases with the increase in $n_c$. $MoSe_2$ waveguides show the highest value of the crosstalk distance due to the highest refractive index of 4.21 at 1550 nm and approximately the same birefringence value as its competitors. Thus, high refractivity and giant anisotropy of van der Waals materials are key factors enabling superior performance

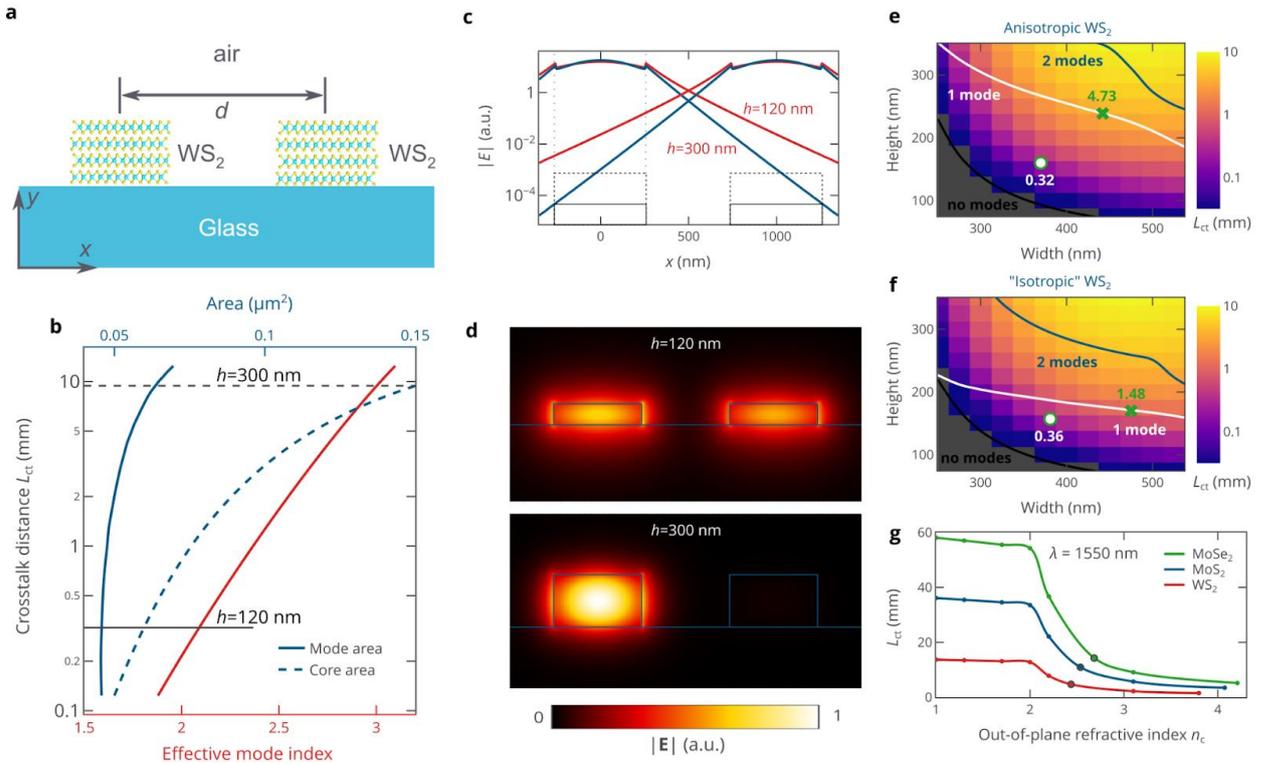

**Figure 2. Theoretical investigation of $WS_2$ waveguides operating at telecom wavelength λ = 1550 nm. a,** Schematic view of a pair of parallel $WS_2$ waveguides, separated by the distance *d*. **b,** Relationship between the crosstalk distance $L_{ct}$, effective mode index, and modal and physical area of 500-nm-wide waveguides at *d* = 1 μm. **c,** Electric field distribution and overlap between the waveguides of the same width w = 500 nm and heights *h* = 120 nm and 300

nm showing the influence of the evanescent tail on the coupling between waveguides and the crosstalk distance. **d,** Distribution of the electric field in the cross-section at a distance $L$ = 0.15 mm from the inlet. At the inlet ($L$ = 0), only the left waveguide carries the signal. **e,f,** Map of the crosstalk distance as a function of the width and height of waveguides separated by the same distance $d$ = 1 μm. In calculations we used: the full anisotropic tensor of dielectric permittivity of $WS_2$ from Figure 1a in panel **e**, and isotropic optical constants with $n_c = n_{ab}$ in panel **f**. The optimal dimensions for a single-mode regime are marked with a green cross. The white circle labels dimensions for which the mode area of the fundamental mode is minimized. **g,** Maximum cross-talk distance in the single-mode regime as a function of the out-of-plane refractive index for different materials. All curves start at $n_c$ = 1 and end at $n_c = n_{ab}$. The circles mark points calculated at actual $n_c$ of the corresponding materials.

## Anisotropic nanowaveguides

To verify our theoretical findings, we fabricated a set of $WS_2$ waveguides (Figure 3a) of the same height of 250 nm, dictated by the thickness of the original $WS_2$ flake, and different widths from 250 to 450 nm in steps of 50 nm. Next, we conducted a study of transverse electric (TE) modes inside fabricated waveguides using s-SNOM in reflection mode (Figures 3d-e) with a large curvature (∼120 nm) probe. It allows exciting horizontal dipole and, hence, measuring TE mode. We excited modes at a standard 1550 nm telecommunication wavelength. In our configuration, the mode is excited at the probe and scatters by a grating. Apart from map scans (Figure 3e), we conducted horizontal line scans (Figures 3f-g), namely, multiple measurements along the center of waveguides. This approach yields better visualization of waveguide mode field propagation (Figure 3g). Afterward, we performed a Fourier analysis (Figure 3h) to determine the effective mode index. Comparing the experimental data with the theoretical predictions in Figure 3i yields a perfect match, further supporting our results and conclusions. Notably, no other modes were detected, thanks to the giant optical anisotropy of $WS_2$. Therefore, optical anisotropy is an up-and-coming resource for efficient light manipulation in next-generation integrated circuits.

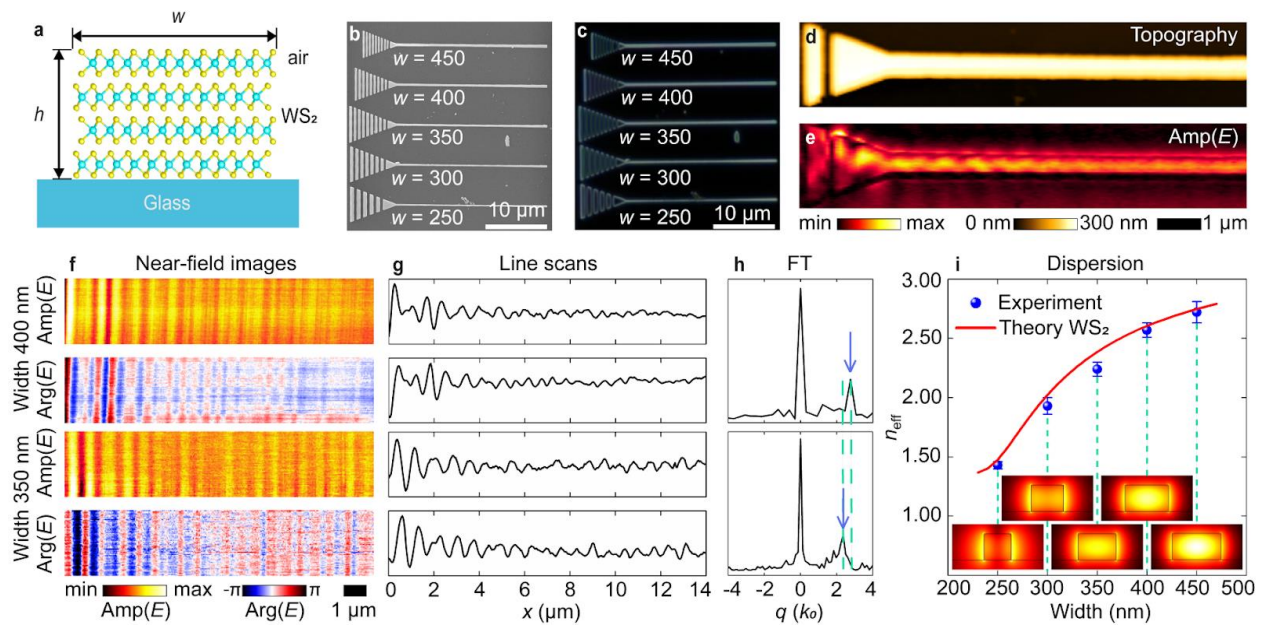

**Figure 3. Experimental investigation of $WS_2$ waveguides. a,** Schematic layout of a single $WS_2$ waveguide. **b,** Scanning electron microscopy and dark-field optical **c,** images of fabricated waveguides. All waveguides have the same height (250 nm), determined by the thickness of the original $WS_2$ flake. **d,** Topography and **e,** field amplitude maps measured by s-SNOM. **f,** Near-field maps of amplitude and phase in 350 and 400-nm-wide $WS_2$ waveguides in the waveguide center. **g,** Line scans and **h,** their Fourier transforms from which we experimentally retrieve effective index of the waveguide mode. **i,** Plot of the measured effective mode index versus the waveguide width and its comparison with the theoretical predictions. The insets show the field distributions of the fundamental guided mode inside waveguides with different widths.

## Discussion

To summarize, we comprehensively studied layered materials' performance in photonic integrated circuitry. First, we measured the anisotropic dielectric tensor of bulk layers of $WS_2$ and verified it via near-field microscopy. As a result, we report the giant optical anisotropy and higher in-plane refractive index than other materials with a similar optical bandgap. Next, we theoretically demonstrated how these unique optical properties can be harnessed to achieve ultra-compact integration in photonic integrated circuits. This superior performance originates from two factors: (i) the explicit advantage of high in-plane refractive index of $WS_2$ and other layered materials, resulting in an unusually high effective index for the fundamental waveguide mode; (ii) hidden advantage of the relatively low out-of-plane refractive index that does not affect the effective mode index of the fundamental mode, yet significantly diminishes effective indices of the higher-order modes. The second factor leads to the increase in the cut-off waveguide core size for higher-order modes. As a result, one is able to maintain the single-mode regime in waveguides with larger cross-sectional area, leading to a higher effective mode index and, eventually, to lower waveguide cross-talk and increased integration density. Finally, we confirmed the excellent waveguiding characteristics of $WS_2$ waveguides by near-field microscopy whose results perfectly agreed with the theory. Thus, our results create a firm ground for applications of layered materials in integrated photonics and establish optical anisotropy as a valuable tool for optical engineers.

## Acknowledgments

L.M.M. acknowledges Project PID2020-115221GB-C41, financed by MCIN/AEI/10.13039/501100011033, and the Aragon Government through Project Q-MAD.

## Author contributions

A.A.V., A.V.A., K.S.N., and V.S.V. suggested and directed the project. G.A.E., D.V.G., and D.I.Y. performed the measurements and analyzed the data. A.A.V., G.A.E., D.V.G., K.V.V., I.K., A.B.M., and I.A.K. provided theoretical support. A.A.V., G.A.E., and D.V.G. wrote the original manuscript. A.A.V., G.A.E., A.V.A., K.S.N., and V.S.V. reviewed and edited the paper. All authors contributed to the discussions and commented on the paper.

## Competing interests

The authors declare no competing interests.

## Methods

**Spectroscopic ellipsometry.** Ellipsometric measurements were performed using the imaging spectroscopic ellipsometer Accurion nanofilm_ep4, which allows collecting the signal from a small spot with a size down to ~10 μm in a broad spectral range from 360 to 1700 nm.

**SNOM.** For near-field measurements we used commercial scattering scanning near-field optical microscope (s-SNOM, www.neaspec.com). As a source we used a monochromatic tunable laser of the wavelength from 1475–1600 nm spectral interval. All the measurements were conducted in the reflection mode (the same parabolic mirror was used to focus and collect the signal from the probe-sample interface) in TE polarization. Tapping frequency of the probe was around 275 kHz, amplitude was near 55 nm. For noise suppression we used a pseudo-heterodyne interferometer at the third harmonic.

**Eigenmode simulations.** To find the eigenmodes of a single and a pair of $WS_2$ waveguides we have employed a finite-element method solver of the Maxwell equations implemented in COMSOL Multiphysics (RF module). Mesh consisted of triangular elements with a characteristic size from 20 nm (at waveguide cores and their proximity) to 250 nm (at the boundary of the simulation domain). The simulation domain had the size of 8 µm × 8 µm. To determine the crosstalk distance, we performed eigenmode simulations for a pair of identical parallel waveguides (Figure 2a). Knowing the effective indices $n_{sym}$ and $n_{asym}$, of symmetric and antisymmetric supermodes, we evaluated the crosstalk distance as:

$$L_{ct} = \frac{\pi}{|\beta_{sym} - \beta_{asym}|} = \frac{\lambda}{2|n_{sym} - n_{asym}|}, \qquad (1)$$

where $\beta$ and $\lambda$ denote the wavevector of the eigenmode and the free space wavelength of light at operating frequency. Mode area was determined as $A_m = \int w_E(x,y)dxdy/\max[w_E(x,y)]$, where $w_E(x,y) = \frac{1}{16\pi}\left(\boldsymbol{E}^*(x,y), \frac{\partial(\omega\varepsilon)}{\partial\omega}\boldsymbol{E}(x,y)\right)$ is the electric energy density and maximum was taken over the waveguide core.

**First-principles calculations.** Optical constants of $WS_2$ ($a = b$ = 3.156 Å, $c$ = 12.35 Å) were calculated within single-shot GW approximation using Vienna Ab-initio Simulation Package (VASP).[42–44] First, the atomic positions of the crystal were relaxed until the interatomic forces were less than $10^{-3}$ eV/Å, and the initial ground state wavefunctions were obtained using standard self-consistent DFT routines. Second, the real and imaginary parts of a frequency-dependent dielectric function were calculated on top of the GW-corrected eigenstates. We used the projector augmented wave pseudopotentials of GW type[45] to describe the behavior of core electrons and their interaction with valence electrons. Cutoff energy for a plane-waves basis was set to 500 eV. Both DFT and GW steps were performed with -centered 12×12×3 k-points grid.

**Sample fabrication.** $WS_2$ flakes were mechanically exfoliated from 2H-phase $WS_2$ crystal (2D Semiconductors Inc., USA) and deposited on $SiO_2$(300 nm)/Si substrate. Waveguides patterning was made by electron beam lithography (EBL) and subsequent reactive ion etching (RIE) of $WS_2$ through the Al hardmask. First, a thin 40 nm aluminum layer was deposited onto the substrate using e-beam evaporation technique (Plassys MEB550S, France). Then, AR-P 6200.04 (from AllResist Gmbh) positive resist was spin-coated at 2000 RPM and soft-baked at 150 °C, resulting in an about 140 nm film thickness. Next, a negative version (with 5–10 µm outside border) of the required waveguides design was exposed with an area dose of 130 uC/cm$^2$ using Crestec CABL9500C EBL system (Japan) at 50 kV accelerating voltage and 350 pA beam current. Patterns in resist were developed using AR 600-546 (AllResist) developer. Next, a RIE of Al layer was carried out using Trion MLIII system (USA) in $BCl_3/Cl_2$ (10:1) gas mixture at 40 mTorr and 100 W RIE applied power. Residual resist was dry-etched by inductively coupled plasma (ICP) etching in oxygen (Corial 200I, France). In order to transfer the obtained pattern from aluminum hard mask to flake, $WS_2$ was anisotropically dry-etched in $C_2H_4/SF_6$ (1:2) gas mixture at 7 mTorr and 20 W RIE with 150 W ICP applied power. Finally, Al layer was removed by ion etching in Plassys MEB550S system using KDC40 DC gridded ion source (KRI, USA) with Ar gas at 5·10$^{-4}$ mbar, 300 V acceleration voltage and 50 mA beam current.

**Data availability**